\documentclass[preprint]{aastex}

\usepackage{emulateapj5}
\usepackage{apjfonts}

\shorttitle{Cosmic Ray Modified Shocks}
\shortauthors{Kang {\it et al.~}}
\slugcomment{draft of \today}

\eqsecnum

\def\eg{{\it e.g.,}}
\def\ie{{\it i.e.~}}

\def\cm3{~{\rm cm^{-3}}}

\def\lsim{\mathrel{  
        \raise0.3ex\hbox{$<$}\kern-0.75em{\lower0.65ex\hbox{$\sim$}}}}
\def\gsim{\mathrel{
        \raise0.3ex\hbox{$>$}\kern-0.75em{\lower0.65ex\hbox{$\sim$}}}}

\begin{document}
\title{Time Evolution of Cosmic-Ray Modified Plane Shocks}

\author{Hyesung Kang}
\affil{Department of Earth Sciences, Pusan National University,
    Pusan 609-735 Korea} 
\email{kang@uju.es.pusan.ac.kr}

\author{T.W. Jones}
\affil{Department of Astronomy, University of Minnesota, Minneapolis, 
      MN 55455}
\email{twj@msi.umn.edu}

\author{R. J. LeVeque}
\affil{Department of Applied Mathematics and Department of Mathematics , University of Washington, Seattle, WA 98195} 
\email{rjl@amath.washington.edu}

\and

\author{K. M. Shyue}
\affil{Mathematics Department, National Taiwan University, Taipei, Taiwan}
\email{shyue@math.ntu.edu.tw}

\altaffiltext{1}{Submitted to the Astrophysical Journal}

\begin{abstract}
We have developed a novel computer code designed to follow the
evolution of cosmic-ray modified shocks, including the full momentum
dependence of the particles for a realistic diffusion coefficient
model. In this form the problem is technically very difficult, because
one needs to cover a wide range of diffusive scales, beginning with
those slightly larger than the physical shock thickness. With most
finite difference schemes for Euler's equations the numerical shock
thickness is at least one zone across, so this provides a lower bound
on the physical scale for diffusive transport computation. Our code
uses sub-zone shock tracking \citep{levshy95} and 
multi-level adaptive mesh refinement \citep{berglev98} 
to provide enhanced spatial resolution around
shocks at modest cost compared to the coarse grid and vastly improved
cost effectiveness compared to a uniform, highly refined grid. We 
present and discuss the implications from our initial results.
\end{abstract}


\section{Introduction}

Diffusive shock acceleration (DSA) is now widely accepted as the model 
to explain the production of cosmic rays (CR) in a wide range of astrophysical 
environments \citep{dru83,blaeic87,berzkry88}.
The concept behind DSA, first-order Fermi acceleration of charged
particles trapped between convergent flows across a shock, is quite
simple. However, the full DSA problem 
is actually extremely complex, because the nonlinear
interactions between energetic particles, resonantly scattering waves and
the underlying plasma can become dominant effects.
Important consequences of nonlinear interactions include such things as
generation and damping of the scattering wave field, injection of suprathermal
particles into the CR population, as well as
heating and compression of the plasma flow due to the CR pressure.
Owing to these complex nonlinear physics involved in the model, 
numerical simulations have been quite useful and successful in understanding 
the details of the acceleration
process and dynamical feedback of the CRs to the underlying plasma
\citep{falgid87,EMP90,dorfi90,kanjon91,berz94,berzvolk00}.

In continuum approaches to numerical simulations of DSA theory, 
the CR diffusion-convection equation is solved at each of a large
number of suprathermal momentum values simultaneously with a set
of fluid equations describing the flow associated with the bulk, thermal
plasma, including the nonlinear interactions between the plasma, CRs and
scattering waves.
Particle acceleration is effected by diffusion across velocity gradients
in the motion of the scattering centers, which are usually assumed to be
tied to the bulk flow. Pressure by the diffusing CRs, in turn, decelerates
and compresses flow into the shock, forming a shock ``precursor''. Since
that development eliminates the original, simple velocity jump seen
by the CRs, the DSA is then modified according to details of the
flow within the precursor, whose scales are characterized by the so-called
diffusion length, $D_{\rm diff}(p) = \kappa(p)/u$, where $\kappa$ is the
spatial diffusion coefficient for CRs of momentum $p$, and $u$ is the 
characteristic flow velocity against which the CRs must swim, \eg \citep{kanjon91}.
Accurate solutions to the CR diffusion-convection equation
require a computational grid spacing significantly smaller than $D_{\rm diff}$,
typically, $\Delta x \sim 0.05 D_{\rm diff}(p)$.
In a realistic diffusion transport model, 
it is thought that the diffusion coefficient 
should have a steep momentum dependence, $\kappa (p) \propto p^s$, 
with $s \sim 1-2$.
For the lowest energy CR particles the diffusion lengths ($D_{\rm diff}(p)$)
are only slightly 
greater than the shock thickness, while they can be many orders of 
magnitude greater than that for the highest energy particles.
Thus, a wide range of length scales is required to be resolved in order 
to solve the diffusion convection equation correctly
for the model with a realistic diffusion coefficient.
Previous numerical simulations which adopted the traditional 
flux-differencing method on a uniform grid were often forced to assume 
a weak momentum dependence, for example,  $s=0.25$ in \citet{kanjon91}. 

To overcome this numerical problem, \citet{berz94} introduced a 
``change of variables technique'' in which the radial coordinate is
transformed into a new variable, $x(p)= \exp[-(r-R_s)/D_{\rm diff}(p)]$ where
$R_s$ is the shock radius, defined for each particle momentum for 
the upstream region. A uniform grid is used for the downstream region. 
This allowed them to solve the coupled system of gasdynamic
equations and the CR transport equation even when the diffusion coefficient  
has a strong momentum dependence (\eg $\kappa(p)\propto p$). 
Their code is designed for simulations of supernova remnants, which are 
represented by piston-driven spherical shocks in one-dimensional geometry.
It is different from conventional Eulerian codes in several ways.
Both gasdynamic equations and the CR transport equation are solved
separately either side of the gas subshock. 
Then the gasdynamic solutions at both sides of the subshock are used 
to solve the Riemann problem, which 
determines how the subshock evolves.
Also an iteration scheme is applied to match the downstream and upstream
solutions for the CR diffusion-convection equation at the subshock.
In any case this has enabled them to explore several important
issues regarding the particle acceleration at supernova remnants 
more fully than was possible before, \eg
\citet{berz95,berz96,berzvolk00}.
However, no consistency-check for this method has been attempted 
so far, since no existing conventional codes can handle such a strongly momentum
dependent diffusion coefficient.

Fermi shock acceleration affects those particles with a mean free path 
greater than the shock thickness that can resonantly scatter with
self-generated Alfv\'en waves.
In the so-called ``thermal leakage'' type injection model,
the diffusion and acceleration of these particles out of the suprathermal 
tail of the Maxwellian distribution determines the CR injection rate
\citep{elleich84,kanjon95}.
A self--consistent, analytic and nonlinear model for ion injection 
based on the interactions of the suprathermal particles 
with self--generated magneto--hydrodynamic waves in strong shocks
has been presented by \citet{mal98}.
By adopting this analytic solution, \citet{gies00} have developed a 
numerical treatment of the injection model at a strong quasi--parallel shock,
which is then incorporated into the combined gas dynamics and the CR 
diffusion--convection code.
Since the suprathermal particles have mean free paths a few times 
that of thermal particles, resolving these smallest scales 
is of critical importance in estimating the injection and acceleration 
efficiency in such numerical simulations of the CR modified shocks.
In fact, \citet{gies00} were able to run their simulations, with a conventional
Eulerian scheme on a uniform grid, only up to the time when the maximum
accelerated momentum was of order of $p_{\rm max}/m_{\rm p}c \sim 1$ for a Bohm
type diffusion model 
because of severe requirements for computational resources needed to
evolve the CR distribution to highly relativistic momenta.
This calls for an alternative method comparable to Berezhko's code, which
solves the CR diffusion-convection equation on a grid whose spacing 
scales with the diffusion length at each momentum value sampled. 

In this contribution, we present a new numerical scheme that follows
CR modified shocks in one dimensional, plane-parallel geometry.
We take advantage of the fact that the diffusion and acceleration 
of the low energy particles are important only close to the shock owing 
to their small diffusion lengths. They are simply advected along with 
the underlying gas flow far upstream and downstream of the shock.
Thus it is necessary to resolve numerically the diffusion length of the 
particles only around the shock. 
So we first implement a shock tracking scheme to locate the shock position
exactly and then increase the grid resolution only around the shock
by applying multi-levels of refined grids.  
Toward this end, we have adopted the shock tracking method of
\citet{levshy95} and the Adaptive Mesh Refinement (AMR) technique 
of \citet{berglev98}, and modified the code to use 
multiple levels of grid refinement only around the shock. 

In the following section we outline our numerical methods,
while in \S 3 we present and discuss our test results. 
Section 4 provides a summary.

\section{Numerical Method}

The diffusive transport model for CR acceleration
separates the plasma into two components distinguished by
scattering length. 
The bulk plasma consists of thermal particles whose scattering 
lengths are small enough to fit within a dissipative shock. 
They are described by the standard gasdynamic equations with CR pressure terms
added \citep{mcv82}.
The diffusion-convection equation, which describes the time evolution 
of the CR distribution function $f(p,x,t)$ (\eg \citet{ski75}) is given by
\begin{equation}
{df\over dt} = {1\over3} (\vec\nabla\cdot \vec u~)  p {\partial f\over
\partial p} + \vec\nabla\cdot (\kappa(x,p) \vec\nabla f), 
\label{diffcon}
\end{equation}
where $d/dt$ is the total time derivative in the fluid frame
and the diffusion coefficient $\kappa(x,p)$ is assumed to be a scalar.
As in our previous studies, the function $g(p)=p^4f(p)$ is solved instead
of $f(p)$.
Except for the special shock tracking and AMR features,
our treatments of the underlying gas dynamics and the CR transport are
relatively standard \citep{kanjon91, gies00}, so we do not repeat them here. 

The spatial diffusion coefficient can be expressed
in terms of a mean scattering length, $\lambda$, as
$\kappa(x,p) = {1\over 3} \lambda v$, where $v$ is the particle speed.
The scattering length, $\lambda$, and thus $\kappa(x,p)$, should be
in principle determined by the intensity of resonantly interacting 
Alfv\'en waves.
For example, the Bohm diffusion model represents a saturated wave spectrum
and gives the minimum diffusion coefficient
as $\kappa_{\rm B} = 1/3 r_{\rm g} v$
when the particles scatter within one gyration radius ($r_{\rm g}$)
due to completely random scatterings off the self-generated waves.
This gives
$\kappa_{\rm B} \propto ~{p^2}/{(p^2+1)^{1/2}}$.
Hereafter we will express particle momenta in units $m_pc$. 
We consider here only the proton CR component.
For our test runs, we will also adopt a power-law form as
$\kappa(p) \propto p^s$ for low momenta ($p<1$) in some models 
in addition to $\kappa_{\rm B}$. 
We note that the Bohm diffusion coefficient becomes $\kappa(p) \propto p^2$ in 
the limit of $p<<1$ and $\kappa(p) \propto p$ in the limit of
$p>>1$.
In order to model amplification of self-generated turbulent waves 
due to compression of the perpendicular component of the magnetic field, 
the spatial dependence of the diffusion is modeled as
\begin{equation}
\kappa(x,p) = \kappa(p)(\rho_1/\rho(x)),
\end{equation}
where $\rho_1$ is the upstream gas density. 
This form is also required to prevent the acoustic instability
of the precursor \citep{drufal86,kanjonryu92}.

We also adopt the thermal leakage type injection model introduced
in \citet{kanjon95}.
In this model, below a certain momentum, $p_1$, chosen high enough to
include most of the postshock thermal population, 
the distribution is forced to
maintain a Maxwellian form consistent with the local gas temperature
and density determined from the gasdynamical equations. 
Above $p_1$ particles are allowed to evolve
according to the diffusion-convection equation, so the form will deviate
from Maxwellian.  However, only for $p \ge p_2 > p_1$
are they included in calculations of CR pressure and energy.
We relate $p_1$ and $p_2$ to the peak of the postshock Maxwellian distribution, 
$p_{th}$, as $p_1 = c_1 p_{th}$, and $p_2 = c_2 p_{th}$,
and we assume $c_1=2.5$ and $c_2=3.0$ for all test runs here. 
Here $p_{th}$ corresponds to the peak in the
partial pressure of thermal particles.
The choice of $p_1$ influences the injection rate directly,
since it determines the fraction of suprathermal particles 
in the Maxwellian tail that can be injected into the CRs.
 
\subsection{Shock Tracking Method}
The hydrodynamic conservation equations are solved in the 1D plane-parallel 
geometry by the {\it wave-propagation} algorithm described in \citet{lev97}.
In this method a nonlinear Riemann problem is solved at each interface between grid 
cells, and the wave solutions (\ie, speeds of waves and jumps associated 
with three wave modes) are used directly to update the dynamic variables 
at each cell. 
Within this method a sub-zone shock-tracking algorithm of \citet{levshy95} can be 
incorporated easily, since the Riemann solutions tell us exactly
how the waves propagate.
The underlying Eulerian grid, which is called the ``base'' grid through this
paper, has uniform cells. 
An additional cell boundary is introduced at the location of the shock, 
subdividing a uniform cell into two sub-cells. 
In the next time step, this cell boundary (shock front) is moved to a new 
location using the Riemann solutions at the current shock location
(\ie, $x_s^{n+1}=x_s^n + v_s~\Delta t$) 
and the waves are propagated onto the new set of grid zones. 
Since the new grid is chosen so that the shock wave coincides exactly 
with an irregular cell boundary, {\it the shock remains as an exact discontinuity 
without smearing}.
One advantage of using the wave-propagation method for the shock tracking 
scheme is that the large time step satisfying the Courant condition for 
the uniform grid can be used even if the shock is very close
to the boundary of the uniform cell and so the sub-cell is much smaller 
than the uniform cell.

The CR diffusion-convection equation is solved in two steps: 1) the diffusion 
term is solved by the Crank-Nicholson scheme as described in \citet{kanjon91}. 
2) the advection term is solved by the wave-propagation method as for the 
gasdynamic variables.

\subsection{Adaptive Mesh Refinement}
Ideal gasdynamic equations in 1D planar geometry do not contain any intrinsic 
length scales to be resolved, but once the precursor due to the CR pressure 
modification becomes significant, 
the grid spacing should be fine enough to resolve the precursor structure.
According to previous numerical studies \eg~\citep{jonkan90,kanjon91}, 
convergence of numerical solutions to CR modified shocks, especially the 
subshock transition, requires that the precursor be resolved with sufficient 
accuracy. 
While the full extent of the precursor increases with the CR pressure and is
related with the diffusion length of the maximum accelerated momentum 
$p_{\rm max}$, the dominant scale length of the precursor is similar to an averaged
diffusion length of the particle populations with the greatest 
contribution to the CR pressure.
Typically a strong shock becomes significantly modified due to nonlinear feedback 
from the CR pressure when the maximum acceleration momentum becomes 
mildly relativistic (\ie $p_{\rm max} \sim 1$).
Thus in order to follow the development of the precursor and the time 
evolution of the CR modified shock,
the gasdynamic equations should be solved on a base underlying grid
whose spacing is smaller than the diffusion length of mildly relativistic
particles.
As discussed earlier, it would be most natural to solve the CR transport 
equation on a grid whose spacing scales with the particle's momentum
as in \citet{berz94}.
In that case, the CR distribution $f(x,p)$ should be mapped onto the base
hydrodynamic grid in order to calculate the CR pressure and its dynamical
feedback on the dynamics of the underlying flow.

\subsubsection{Laying down multi-level grids}
Here we take a different approach in which the immediate upstream and downstream
regions around the shock are refined by applying multi-level grids with
increasingly finer resolution by an integer factor of two, so that the 
transport of low energy particles right adjacent to the shock is at the 
most refined grid.
Here we refer each level grid by the grid level index $l_g$ which runs from 
0 to $l_{\rm max}$, corresponding to the base grid and the finest grid,
respectively.
Since the grid spacing decreases by an integer factor, we can lay down
the refined grid in such a way that cell boundaries align between 
two adjacent levels. This feature allows us to use a much simpler 
mapping scheme between two adjacent levels, compared to the case where 
non-integer refinement factors are used.  
We adopt the Adaptive Mesh Refinement technique developed by 
\citet{berglev98}.
In the general version of the AMR code of \citet{berglev98}, the code identifies 
the refinement regions where the desired level of numerical accuracy is not 
achieved and the multi-level grids are generated within the refinement
regions. Compared to that general version, a much simpler scheme is 
sufficient for our needs, since we only need to refine the region around the 
shock whose location is exactly known in our shock-tracking code.
In fact, it is crucial to have the shock-tracking in order to lay down the
multi-level grids around the shock, so that the shock remains near the 
middle of the computational domain at all levels.

A fixed number of cells around the shock ($N_{\rm rf}$) are identified as the
``refinement region'' on the base grid (\ie $l_g= 0$ grid).
The 1st level grid is generated by placing $2N_{\rm rf}$ cells within the
refinement region, so each cell is refined by a factor of two. 
We use the refinement factor of two, since it is relatively simpler in
terms of programming and it improves robustness and stability of the code.
Then $N_{\rm rf}$ cells around the shock out of $2N_{\rm rf}$ cells on the 
1st level grid are chosen to be refined further to the 2nd level grid, 
making the length of the refinement region a half of that in the 1st 
level grid.
Here the refinement region is chosen so that the shock is always in its
middle. 
The same refinement procedure is applied to higher level grids.  
So at all levels, there are $2N_{\rm rf}$ cells around the shock, but the
length of the computation domain is shrunk by a factor of 2 from the 
previous level.
At each level the grid spacing is given by 
$\Delta x(l_g) = \Delta x(0)/2^{l_g}$, where $\Delta x(0)$ is the grid
spacing at the base grid.
Fig. 1 shows an example of refined grid levels up to $l_{\rm max}=2$ 
with $N_{\rm rf}=4$. The value of $N_{\rm rf}$ should be chosen so that
the refined region at the base grid includes most of the precursor during
the early stages when particles get accelerated to mildly relativistic energies.
In real test simulations presented here, typical values of $N_{rf}=100-200$.
The spatial extent of the highest refined grid can be much smaller than the 
length scale of the precursor, since it need only resolve structures
``seen'' by the lowest energy CRs immediately next to the shock.  

In order to ensure that the shock remains near the middle of the computational 
domain at all grid levels during the time integration of one time step of the 
base grid, we do the following procedure.
First, the velocity in the refined grid is transformed so that the shock 
is at rest in the frame of the numerical simulation at each level 
except the base grid.
Secondly, the multi-level grids are redefined at each time step so that 
they center around the new location of the shock, and all hydrodynamic 
variables and the particle distribution function $g(p)$ are mapped 
onto the newly defined grids.  
Thus the refinement region at all levels is moving along with the shock. 
Although the original shock tracking code of \citet{levshy95} can treat 
multiple shocks, we modify it in the current version of our code
to include only one shock, in order to keep the code structures as simple
as possible for our initial studies.

\subsubsection{Time integration}
Integration of the gasdynamic variables and advection of the CR distribution 
function are done by the wave-propagation method at each level grid.
The time step at each level, $\Delta t(l_g)$, is determined by a standard 
Courant condition, that is, $\Delta t(l_g) = 0.3 \Delta x(l_g) / \max
( u + c_s)$, where $u$ and $c_s$ are the flow velocity and sound speed 
at each cell, respectively.
If the highest level is specified to be $l_{\rm max}=1$, 
to advance from $t^{n}$ to $t^{n+1}= t^n + \Delta t(0)$ at the base grid,
we need the following steps:
1) all equations are integrated at the base grid.
2) the same is done twice with $\Delta t(1) = \Delta t(0)/2$
at the $l_g = 1$ level for the refined region.
The cells immediately outside the refinement region at the $l_g=0$ level
provide the necessary boundary conditions for the integrations.
Here we need boundary conditions at two spatial points at both ends of the
refined grid, $q_0$, $q_1$, $q_{2N_{rf}}$, and $q_{2N_{rf}+1}$, 
and also at two points in time, $t^n$ and $t^{n+1/2}$.
Some of them are interpolated in time and space coordinates
from the variables defined at the grid one level below.
3) the values at the $l_g=1$ level are mapped onto the refinement region 
at the base grid.
4) finally the values at the interface just outside the refinement region
at the base grid  should be corrected to preserve global conservation.  
The base idea and also the detailed procedure applied for one level of 
refinement can be found in \citet{berglev98}.
When $l_{\rm max}$ is greater than 1, the same procedures should be repeated 
recursively at each time step at each level. 
The variables within the refinement region at the $l_g$ level grid
are replace with the more accurate values at the $l_g+1$ level grid after
the corresponding pair of time steps at the $l_g+1$ level are completed.
In one time step in the base grid, we carry out $2^{l_g}$ time steps at
the $l_g$ level grid,
so total number of time steps in all level grids required
to advance one time step in the base grid 
becomes the sum of $ 2 + 2^2 + ...+ 2^{l_{\rm max}}$. 

Once advection of the CR distribution is done, diffusive transport including
the first-order acceleration is solved in a separate step. 
Although an implicit Crank-Nicholson scheme is used for the diffusion term, 
the time step is still restricted by the acceleration term as 
$ \Delta t_{\rm CN}(l_g) = 0.5 \min[3 \Delta (\log p)/ (\partial u/ \partial x)_i ]$.
Within a single hydrodynamic time step, several Crank-Nicholson time steps
are performed if $\Delta t(l_g) > \Delta t_{\rm CN}(l_g)$.
Number of this sub-cycling is about 4-5 for the momentum bin size considered
here (\ie, $\Delta (\log p)=0.026$).

As we go up and down on the ladder of ``time stepping'' for one time
step in the base grid, the base grid being the lowest,
the values at a coarser grid propagate upward as boundary conditions
for a finer grid and more accurate values at the finger grid propagate
downward by being mapped onto the coarser grid. 
\section{Test Results}

In this section we present some test simulations with our CR/AMR code.
The dynamics of the CR modified shock depends on four parameters:
the gas adiabatic index, $\gamma_{\rm g}=5/3$, 
gas Mach number of the shock, $M=V_s/c_s$, 
$\beta=V_s/c$, and the diffusion coefficient,
where $c_s$ and $c$ are the upstream sound speed and the speed of light,
respectively.
For all simulations we present here, $\beta=10^{-2}$, and $\gamma_{\rm g}=5/3$.
We considered three values for Mach number, $M=5,10,$ and 20 for the initial
shock jump by adjusting the preshock gas pressure.
The initial jump conditions in the rest frame of the shock for all 
test problems are:
$\rho_1=1., P_{g,1}= 1.5\times 10^{-3} (20/M)^2 , u_1= -1.$ in upstream 
region and $\rho_2=4., P_{g,2}= 7.5\times 10^{-1}, u_2= -0.25$ in 
downstream region.
Here the velocities are normalized to the initial shock speed, 
$V_s=3000{\rm kms^{-1}}$.
Normalization of the length and the time variables depends on the diffusion
coefficient: $\kappa(p) = \kappa_{\rm phys} / \kappa_o$, where
$\kappa(p)$ is the computational coefficient, $\kappa_{\rm phys}$ is
the physical value, and $\kappa_o$ is the normalization constant.
So the corresponding normalization constants are: $t_o = \kappa_o/V_s^2$,
$r_o= \kappa_o/V_s$. These correspond to roughly the diffusion length and
diffusion time scale for $p \sim 1$.
For the Bohm diffusion coefficient with the magnetic field of 1 microgauss,
for example, $\kappa_{\rm B} = ~{p^2}/{(p^2+1)^{1/2}}$ with 
$\kappa_o=3.13\times 10^{22} {\rm cm^2~s^{-1}}$, so that
$t_o=3.5\times 10^{5} {\rm s}$ and $r_o=1.05\times 10^{14} {\rm cm}$.
The particle number density, $n_o$ is arbitrary, but the gas density 
and pressure are normalized to $\rho_o= m_p n_o$ and $P_o = \rho_o V_s^2$,
respectively.
 
We use 230 logarithmic momentum zones in $\log (p)$=[-3.0,+3.0] and the momentum
is in units of $m_{\rm p}c$.
The distribution function $f(p)$ is expressed in units of 
$f_o= n_o/(m_{\rm p}c)^3$, so that $4\pi \int f~p^2~dp = \rho$.

\subsection{Test of Refinement}

In this section we consider the $M=20$ shock with 
a diffusion model $\kappa(p) = p $.
In order to see how the CR/AMR code performs at different 
resolutions, we ran the simulations with different levels of
refinements, $l_{\rm max}= 0, 1, 2, 3, 4$.
The numerical domain is {[-25,+75]} and the number of cells, $N=2000$,
so the grid spacing is $\Delta x(0) =0.05$ which corresponds to 1/20 of the
diffusion length of the particles of $p=1$.
The number of refined cells around the shock is $N_{rf}=200$ on the base 
grid and so there are $2N_{rf}=400$ cells on each refined level.
Since the diffusion length of $p_1\sim 0.01$ is $D_{\rm diff}=0.01$, 
the transport of the suprathermal particles can be resolved at the 
$l_{\rm max} = 4$ level.

With the refinement of $l_{\rm max} = 4$ level, the CR transport for
the particles at the injection pool should be resolved and so the
evolution of the CR should be converged.
Fig. 2 shows the time evolution of the model shock with the maximum
refinement of $l_{\rm max}=4$. 
This shows how the precursor develops and modifies the shock structure
as the CR pressure increases in time. 
The numerical frame is chosen so that the initial shock moves to the 
right with $u_s=0.05$, 
but the simulated shock drifts to the left due to the CR pressure .
Fig. 3 shows the density structure at each refined grid (solid lines) 
for $l_g=1,2,3,$ and 4 levels, superposed on the density structure
at the base grid (dotted lines) in the $l_{\rm max}=4$ simulation
shown in Fig. 2. 
This demonstrates how the size of refinement region decreases
at higher levels and how the refinement regions move along with the shock.
As indicated by two points at downstream and upstream of the subshock
in the top-left panel, 
the shock is tracked as a perfect discontinuity.  

Given the same resolution at the base grid, the simulations with 
larger refinement levels show faster acceleration and faster growth of $P_c$.
Fig. 4 shows how the particle distribution ($g(p)= f(p)~p^4$) at the shock 
evolves with time in the simulations with different refinement levels
for the same shock model shown in Fig. 2:  
there are 5 curves corresponding to $l_{\rm max} =$ 0, 1, 2, 3, and 4
in each panel.
They show the typical Maxwellian distribution that peaks at 
$p_{th}\sim10^{-2.3}$ 
and the CR distribution that asymptotes to a power-law as time increases. 
For $l_{\rm max}=0$ (dotted lines), the cell size $\Delta x=0.05$ is 
too large for the diffusion of the particles in the injection pool to be 
treated correctly, so the injection and the acceleration are under-estimated
and the slope just above the injection pool is steeper than the 
canonical strong-shock test-particle spectrum of $f (p) \sim p^{-4}$. 
For the highest refinement case, $l_{\rm max}=4$ (solid lines), 
the CR pressure becomes dominant over the gas pressure and the compression 
ratio at the subshock becomes 3.3.
As a result, the distribution function steepens from a power-law of $p^{-4}$
at lower momentum just above the injection momentum. 
But it flattens at higher momenta since high energy particles diffuse on
a much larger scale and sample a larger velocity jump. 
These opposite trends lead to a concave curve in the middle, which
is a typical signature of nonlinear effects \citep{berzell99}.
Fig. 5 shows how the CR pressure increases with time in the simulations
with different refinement levels shown in Fig. 4. 
It demonstrates that injection and acceleration are much slower in
under-resolved simulations ($l_{\rm max} <4$). 
From comparison between results of $l_{\rm max} = 3$ (dashed line) and 
$l_{\rm max} = 4$ (solid lines),
we have concluded that the simulation is mostly converged for 
$l_{\rm max} = 4$.
For example, the CR pressure at the shock at $t=$ 50 and 60 is the same
for $l_{\rm max} = 3$ and 4 simulations, although the shock position is
slightly different in the two cases due to slightly different evolution 
in early stages.

The required computing time increases with $l_{\rm max}$ and $N_{rf}$, given
the same resolution at the base grid.
For the simulation considered here in which $N_{rf}/N=0.1$,
the computing time increases by factors of 1.5, 2.3, 4, 7 for the maximum
refinement levels $l_{\rm max} =$ 1, 2, 3, 4, respectively, compared with the
case of no refinement ($l_{\rm max}=0$).  
The computing time would increase by factors of $ (2^{l_{\rm max}})^2$,  
if the simulations were done on a uniform grid spacing that
matches the cell size at the highest refined level grid. 
Fig. 6 compares the computing time for these two cases.
Since only the precursor region needs to be refined, our CR/AMR code
will be most cost-effective for simulations where the precursor is only
a small fraction of the total computational domain.

\subsection{Convergence Test}
In this section we explore how 
a simulation with our CR/AMR code using multi-level refinements would be compared 
with that with no refinement but on a single uniform grid of the same spacing 
as the highest level of the other.  
For this test, we use the following diffusion model: 
$\kappa_{\rm l} = p/\sqrt{2}$ for $p < 1$ and
$\kappa_{\rm h}= p^2/(p^2+1)^{1/2}$ for $p \ge 1$.
This model has a Bohm-type diffusion at higher momenta, but much
weaker momentum dependence at lower momenta.
This allows us to use larger grid spacing near the injection pool 
($p\sim 0.01$)
compared with a Bohm diffusion model which scales as $p^2$ for $p<<1$. 
The initial shock parameters are the same as the $M=20$ shock model 
in previous section.
We consider the following three models:
Model A, a uniform grid with $\Delta x(0)=0.1$ and $l_{\rm max}=0$, 
Model B, an adaptively refined grid  with $\Delta x(0)=0.1$ and $l_{\rm max}=5$,
and Model C, a uniform grid with $\Delta x(0)=0.1/32$ and $l_{\rm max}=0$.
The numerical domain is {[-50,+50]} and the number of cells, $N=1000$
for Models A and B, and $N=32000$ for Model C.
So Model C has the same grid spacing as the $l_{\rm max}=5$ level grid 
of Model B.
The computing time for Model C with a single fine grid is about 150
times longer than that for Model B with the refinement. 
We expect that in the Model A simulation the transport of low energy particles
are under-resolved, so it is included only for comparison.
We choose $N_{rf}=100$ for Model B, so 10 \% of the base grid is refined. 
The major difference between Models B and C is that
the refined grid at $l_{\rm max}=5$ covers only about 1/32 of 
the precursor in Model B, while the entire grid of Model C has the 
finest resolution.

Fig. 7 shows the evolution of the gas density and the CR pressure,
and the CR distribution function at the shock in Models B and C. 
In Model B with refinement the CRs are accelerated a little faster
than in Model C with a single fine grid, so that the difference in the
CR pressure at the shocks is 8.3 \% $t=20$, but this fraction 
decreases to 4.6 \% at $t=40$ and to 3.0 \% at $t=60$. 
Thus, the two methods appear to converge at slightly different rates, but
both give reasonable results once the resolution next to the shock is refined
to resolve $D_{\rm diff}(p_1)$ by more than an order of magnitude.
This test convinces us that our CR/AMR code can perform the intended
simulations with a reasonable accuracy in a very cost-effective way.

\subsection{Dependence on Diffusion Model}
Finally in this section we explore briefly how different diffusion models affect
evolution of the injection and the acceleration efficiency in CR modified shocks.
First we consider the $M=20$ shock and the following three diffusion models.
For lower momenta, $p<1$,
Model K1 uses $\kappa_{\rm l,1} = p/\sqrt{2}$, 
Model K2 $\kappa_{\rm l,2} = p^{1.5}/\sqrt{2}$, and 
Model K3 $\kappa_{\rm l,3} = p^2/(p^2+1)^{1/2}$. 
All three models use the Bohm type diffusion at higher momenta, that is,
$\kappa_{\rm h}= p^2/(p^2+1)^{1/2}$ for $p \ge 1$ and so $\kappa{_{\rm l}}$
continuously matches onto $\kappa{_{\rm h}}$ at $p=1$.
The grid spacing for the base grid, $\Delta x(0)= 0.05, 0.025$, and 0.005
for Models K1, K2, and K3, respectively.
The maximum refinement level, $l_{\rm max}=4,7,$ and 7 for Models K1, K2,
and K3, respectively. 
These parameters are chosen so that the grid spacing at the finest grid is
fine enough to treat the low momentum particles near $p_1$ with
the assumed diffusion model.
While Model K1 was integrated for $t=100$ and Model K2 for $t=60$, Model K3
is integrated only up to $t=20$ due to much longer required computing time.

Fig. 8 shows the gas density, the CR pressure, and the CR distribution
function at the shock for the $M=20$ shock 
at $t=20$ simulated with the three different $\kappa(p)$ 
(K1: solid line, K2: dotted line, K3: dashed line).
Model K3 has the smallest $D_{\rm diff}(p)$, while Model K1 has the largest
$D_{\rm diff}(p)$ for the particles in the injection pool ($p\sim 0.01$).
So the injection takes place the fastest and the CR pressure increases
most efficiently in Model K3 during the early evolution.
According to earlier evolution (not shown), the time evolution of these
three models differs significantly when the maximum accelerated momentum
is still nonrelativistic ($p_{\rm max}<<1$). 
At $t=20$, however, $p_{\rm max}\sim 2$ and mildly relativistic 
particles dominate the CR pressure, so the different diffusion models 
at nonrelativistic momenta no longer play a significant role. 
At this time all three models evolve in a similar way, since the same 
$\kappa_{\rm h}$ is used for all models when $p > 1$. 
Especially Models K2 and K3 show very similar evolution up to this time.
Subsequent development of the shocks is almost independent of the form for
$\kappa_{l}$.
The bottom right panel shows the CR distribution at the shock at $t=60$
in Models K1 and K2 (K3 was ended at $t=20$), 
demonstrating the very similar CR distribution evolution
in Models K1 and K2 at later times.
Considering the earlier trend that Models K2 and K3 are already very similar
at $t=20$, we can deduce all three models evolve the same way when
$p_{\rm max}$ becomes much larger than one.
This implies that $\kappa \propto p$ can be used instead of $\kappa_{\rm B}$
as long as the detailed evolution at early stage when the particles are
still mostly nonrelativistic is not taken seriously.
Simulations with $\kappa(p) \propto p$ model allow much coarser grid
spacings to follow the lowest momentum particles than those with a Bohm type 
diffusion, which reduces the required level of refinements and the 
associated costs. 
The same set of simulations were repeated for $M=10$ and $M=5$ shocks,
and we came to the same conclusion (see Fig. 9 for results from the Mach 10
simulation). 
This validates the notion that $\kappa(p)\propto p$ can be used instead
of $\kappa_B$ in the CR/hydro simulation of SNRs \citep{berz95,berzvolk00}.

We also ran a model with a pure power-law type diffusion model, that is,
$\kappa=p$ for all momenta until $t=10^4$ to study the long-term evolution.
For this simulation, the computational domain is increased to
[-4000,+4000], the zone number in the base grid to $N=40000$ and 
the maximum refinement level to $l_{\rm max}=6$.
Fig. 10 shows the shock structure, the CR distribution at the shock,
and its power-law slope, $ q = -(\partial \ln{f} / \partial \ln{p}) $, 
at $t=$ 200, 800, $2\times 10^3$, $5\times 10^3$, and $10^4$.
The CR pressure at the shock seems to have reached a quasi-steady state value
at $t>200$, even though the maximum momentum continues to increase with time.
The compression ratio at the subshock is $r_{\rm s}\sim 3.1$ at $t=10^3$
which leads to the test-particle slope of $q\sim 4.5$. 
So the particle distribution near the injection pool reflects this slope. 
Although the CR pressure at the shock stays more or less constant after 
$t=200$, and the total shock jump does also, the precursor 
and its associated velocity structure broadens with time.
Thus, since $\kappa(p)$ does not change over time in this
idealized simulation, the particles of a given momentum sample a smaller 
velocity jump, $\Delta u(p)$, around the shock, as the precursor broadens, 
resulting in slightly steeper slope. 
On the other hand, the highest momentum particles sample something close to the
full velocity jump on their diffusion scales, so the slope flattens gradually to
$q\sim 3.2$ toward $p_{\rm max}$, which corresponds to the total compression
ratio, $r_{\rm tot}\sim 11-12$.
This hardening of the CR distribution from $q\sim 4.5$ to $q\sim 3.2$ 
produces concave curves in $\log(g=fp^4)$ versus $\log(p)$ plot.  
This illustrates the importance of following correctly the non-linear 
feedbacks between the CRs and the dynamics inside the precursor, which
requires one to resolve numerically all relevant scales.  

Finally, Fig. 11 shows how the postshock gas temperature decreases as
the CR pressure becomes dominant in the precursor and
how the injection parameter settles down to a constant value at
$\eta \sim 6\times 10^{-4}$ after the shock has reached a quasi-steady state. 
The injection parameter $\eta$ is defined in \citet{kanjon95} and
represents the fraction of the incident proton flux that is injected 
into the CR population at the shock.
We also plot the adiabatic index of the CR population at the shock
which also settles down to a constant value at $\gamma_c\sim 1.37$.
The injection parameter and the CR adiabatic index along with the mean
diffusion coefficient are three free parameters for the so-called two-fluid
model for the CR modified shock simulations.

\section{Summary}

We have developed a new hydro/CR dynamics code which is specifically
designed to solve the time dependent evolution of CR shocks.
Diffusive shock acceleration of the CR particles depends on 
the diffusion of particles whose momenta span many orders of magnitude. 
Since the length and time scales for evolution of the CR kinetic equation 
scale directly with the diffusion coefficient,
an accurate solution to the problem requires that one include all of those
scales in the simulation, beginning just outside the gas subshock thickness. 
Thus, in order to follow accurately the evolution of a CR modified shock,
it is necessary to resolve the precursor structure upstream of the subshock
and, at the same time, to solve correctly the diffusion of the low energy 
particles near the injection pool.
These low energy particles have diffusion lengths that are much smaller 
than the scale length of the precursor, so a large dynamic range
of resolved scales is required for CR shock simulations. 
To solve this problem generally we have successfully combined a
powerful ``Adaptive Mesh Refinement'' (AMR) technique
\citep{bergcol89,berglev98} and a ``shock tracking'' 
technique \citep{levshy95}, and implemented them into a hydro/CR code
based on the wave-propagation method \citep{lev97}. 
The AMR technique allows us to ``zoom in'' inside the precursor structure
with a hierarchy of small, refined grid levels applied around the shock.
The shock tracking technique tracks hydrodynamical shocks and maintains them
as true discontinuities, thus allowing us to refine the region around
the shock at an arbitrary level. 
The result is an enormous savings in both computational time and 
data storage over what would be required to solve the problem using 
more traditional methods on a single fine grid.

The code has been applied to simulations of CR modified shocks with
several diffusion coefficient models with strong momentum dependence,
which were not possible previously due to severe computational requirements.
The main conclusions from the simulations can be summarized as follows:

\begin{enumerate}

\item{}
Our CR/AMR technique code proves to be very cost effective.
In typical simulations where 10\% of the base grid is refined with $l_{\rm
max}$ levels, for example,
the computing time increases by factors of $(2^{l_{\rm max}})^{0.7}$ 
compared with the case of no refinement ($l_{\rm max}=0$).  
It should be compared with the time increases by factors of 
$ (2^{l_{\rm max}})^2$ for the simulations of an uniform grid spacing that
matches the cell size at the $l_{\rm max}-th$ refined level grid.
In a simulation where the precursor scale is only a small fraction of
the computational domain, the advantage in computing time of the refined
multi-level grid over a single fine grid becomes even greater.

\item{}
A convergence test is performed for a Mach 20 gas shock with $V_s/c=0.01$,
which evolves into a CR pressure dominated shock. 
Comparison between a simulation on a coarse grid with multi-level
refinement and another simulation on a single fine grid
without refinement has demonstrated that our CR/AMR code can perform the
intended simulation with reasonable accuracy at a much shorter computing
time. The difference in the CR pressure in two test simulations is around
10\% in early evolution, but deceases to a few \% after the shock has
reached a quasi-steady state in later evolution. The required computing
time is reduced by a factor of 150 in the AMR simulation. 

\item{}
We also carried out a set of simulations when three different diffusion 
models, $\kappa_{\rm l}= p/\sqrt{2}$, $p^{1.5}/\sqrt{2}$, and 
$p^2/(p^2+1)^{1/2}$
for $p<1$, are included, while a Bohm type, $\kappa_{\rm h}= p^2/(p^2+1)^{1/2}$
is assumed for $p\ge1$.
Three simulations generate similar results once the CR pressure is dominated
by the relativistic particles ($p>1$), when the maximum acceleration momentum
becomes $p_{\rm max} >>1$.  Thus a diffusion model of $\kappa \propto p$ can
be used instead of a Bohm model as long as one does not focus on the 
early evolution when $p_{\rm max}<1$. 
Since we can use much larger grid spacings in simulations with 
$\kappa \propto p$ model than those with a Bohm type diffusion at low 
momenta $p<<1$, 
the required level of refinements and so the computing resources
can be reduced greatly. 

\item{}
For a Mach 20 shock, with an injection rate of $\eta \sim 6\times 10^{-4}$,
the shock becomes CR dominated and develops a significant precursor.
Since the flow is decelerated gradually through the precursor,
the velocity jump that the CR particles sample across the shock
depends on the diffusion length of the particle, that is, 
$\Delta u(p) = fcn (\kappa(p))$.
So the slope of the particle distribution function, defined as 
$q(p)=-(\partial \ln{f}/\partial\ln{p})$, increases with $p$. 
In the simulated shock, the compression ratios across the subshock and
across the total transition are 3.1 and 11, respectively, so 
$f(p)$ is $\propto p^{-4.5}$ at low energy momenta but flattens
to $ f(p) \propto p^{-3.3}$ at high energy momenta just below $p_{\rm max}$.
This demonstrates that nonlinear feedbacks between the precursor dynamics
and the CR injection and acceleration should be treated accurately in
numerical simulations of CR shocks.

\end{enumerate}

We are currently implementing a numerical scheme for the
self-consistent injection model by \citet{gies00} which is based on
the plasma-physics study of the nonlinear wave-particle interactions
in shocks presented by \citet{mal98}. 
This will allow us to eliminate any free parameters for the injection 
process from the CR shock simulations.
We intend also to extend the code to treat spherical shocks in order to
study CR acceleration in supernova remnants.

\acknowledgments

HK was supported by Korea Research Foundation Grant (KRF-99-015-DI0114).
TWJ is supported by the University of Minnesota 
Supercomputing Institute, by NSF grant AST-9619438 and by NASA grant 
NAG5-5055.  RJL is supported in part by
DOE grant DE-FG03-96ER25292 and NSF grant DMS-9626645. 
KMS was supported in part by National Science Council of Republic
of China Grant NSC-89-2115-M-002-019.

\clearpage

\begin{figure}
\epsfxsize=14truecm
\centerline{\epsfbox{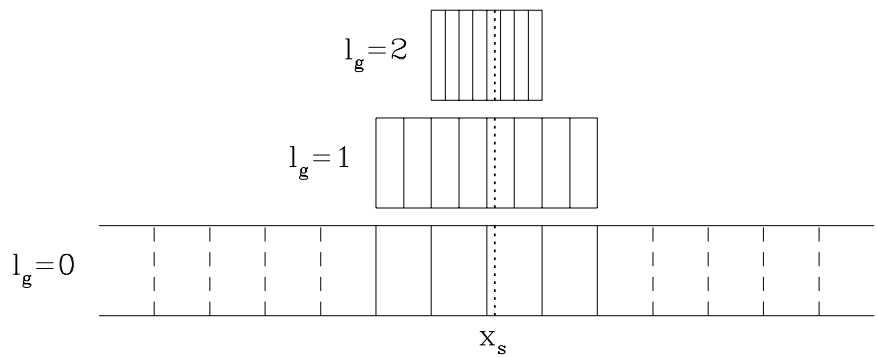}}
\figcaption{
Layout of the base grid and two refined grids. 
Here $N_{rf}=4$ cells around the shock are refined by a factor of two. 
The shock is indicated by the dotted lines. 
\label{fig1}}
\end{figure}

\begin{figure}
\epsfxsize=14truecm
\centerline{\epsfbox{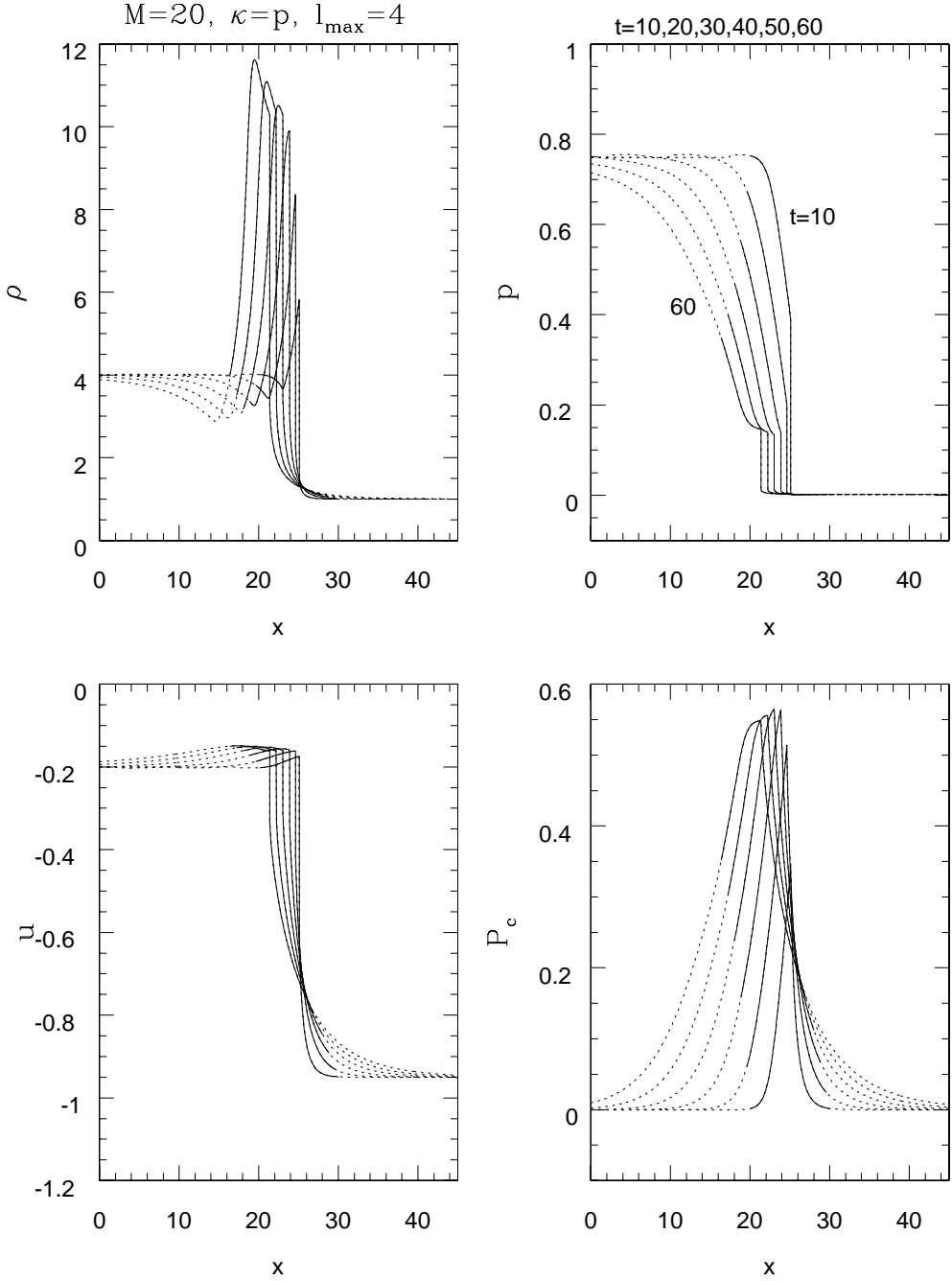}}
\figcaption{
Time evolution of the $M=20$ shock with $l_{\rm max}=4$
refined grid levels at $t=10,$ 20, 30, 40, 50, and 60.
The shock moves to the left, so the right most plots correspond to the 
earliest time $t=10$. 
The solid lines are for the refinement region at $l_g=1$ grid, while the
dotted lines represent the shock structure on the base grid.
\label{fig2}}
\end{figure}

\begin{figure}
\epsfxsize=14truecm
\centerline{\epsfbox{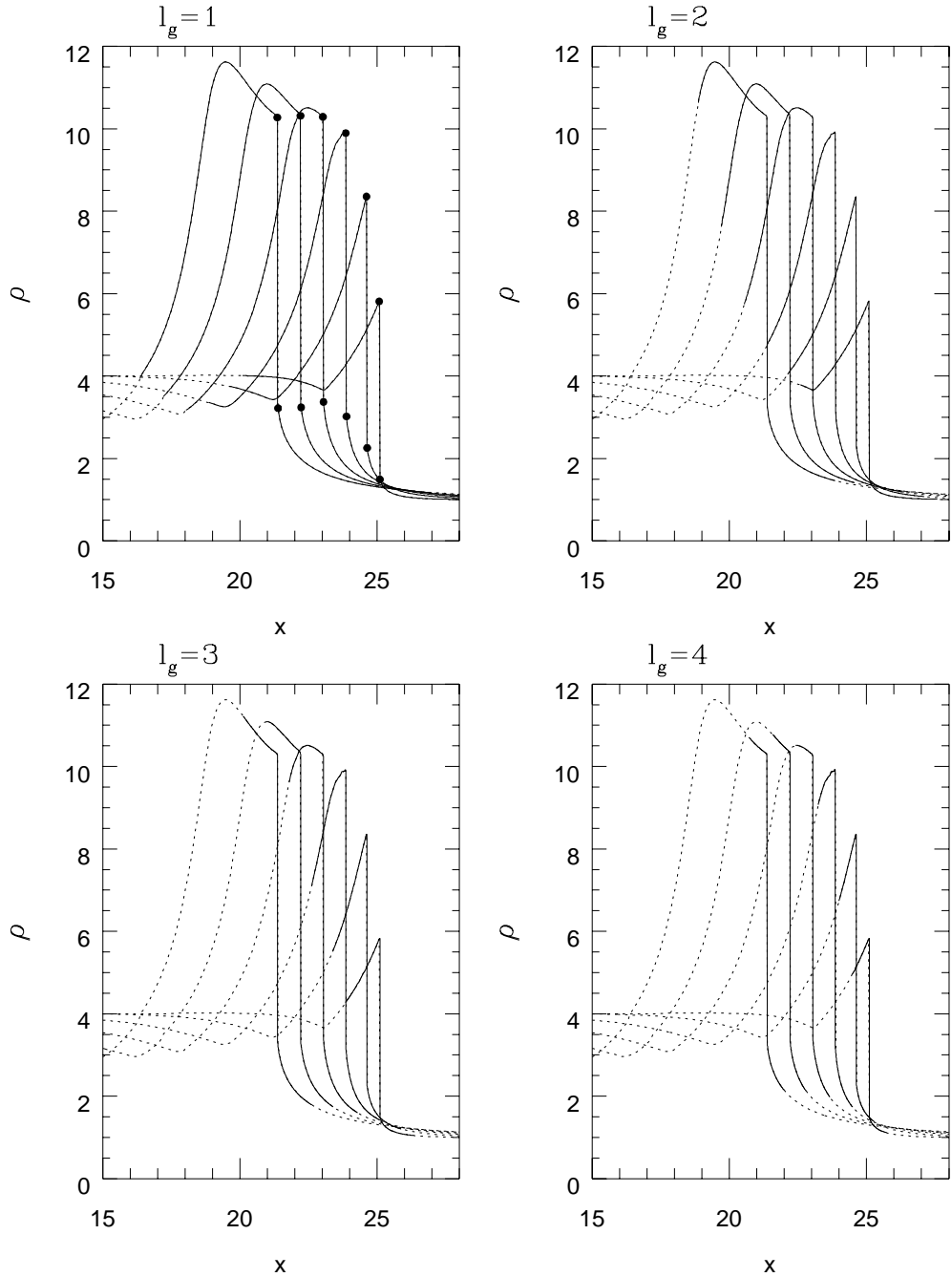}}
\figcaption{
Time evolution of the gas density for the $M=20$ shock in the
simulation with $l_{\rm max}=4$.
The solid lines are for the refinement region at each grid level, while the
dotted lines represent the density profile on the base grid.  
\label{fig3}}
\end{figure}

\begin{figure}
\epsfxsize=14truecm
\centerline{\epsfbox{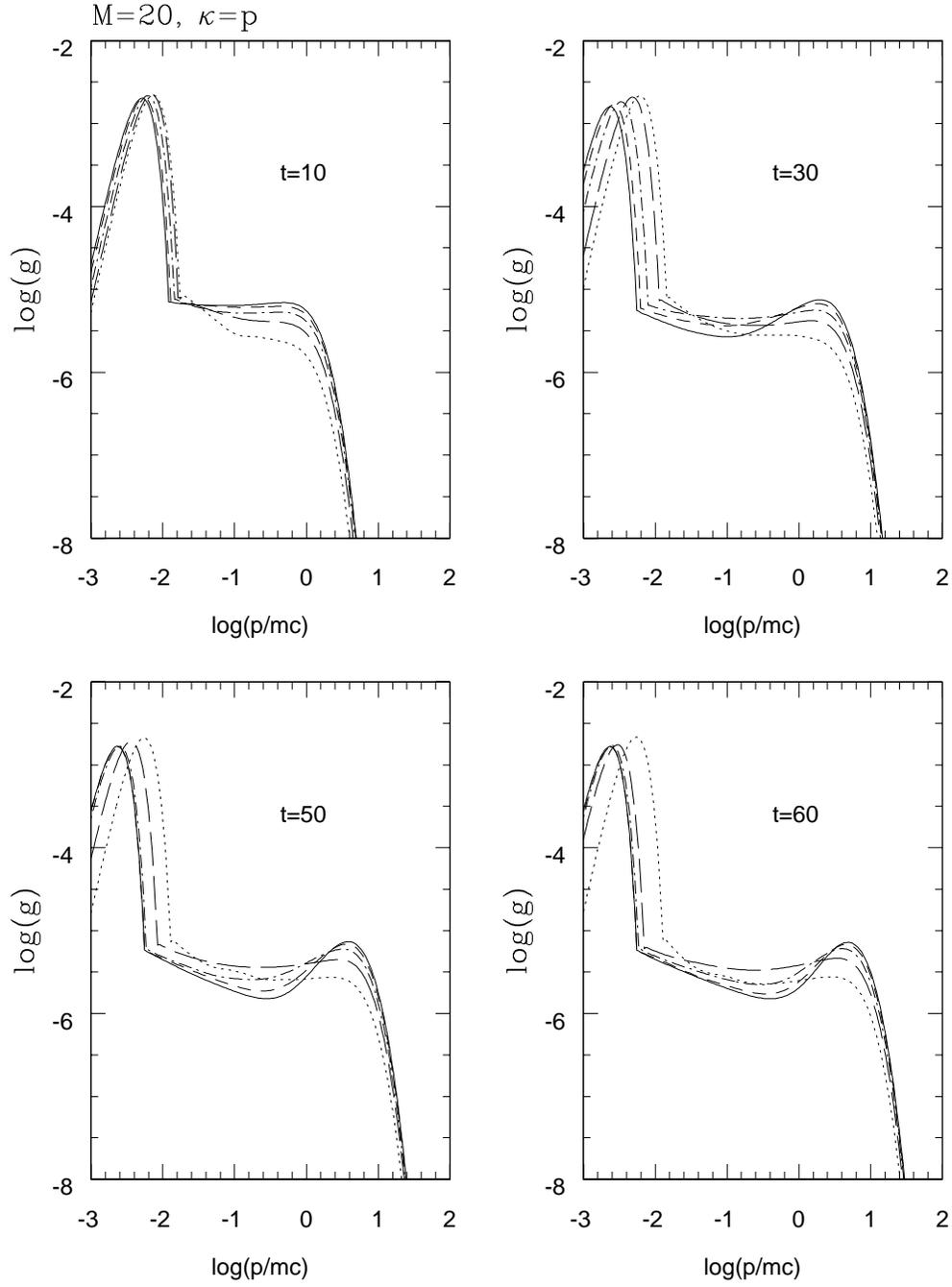}}
\figcaption{
Distribution function ($g=fp^4$) at the shock for the $M=20$ shock 
at $t=10$, 30, 50, and 60. 
The results of the simulations with the maximum refined grid level 
$l_{\rm max}=$ 0 (dotted lines), 1 (long dashed), 2 (dot-dashed), 
3 (dashed)  and 4 (solid) are plotted for comparison.
\label{fig4}}
\end{figure}

\begin{figure}
\epsfxsize=14truecm
\centerline{\epsfbox{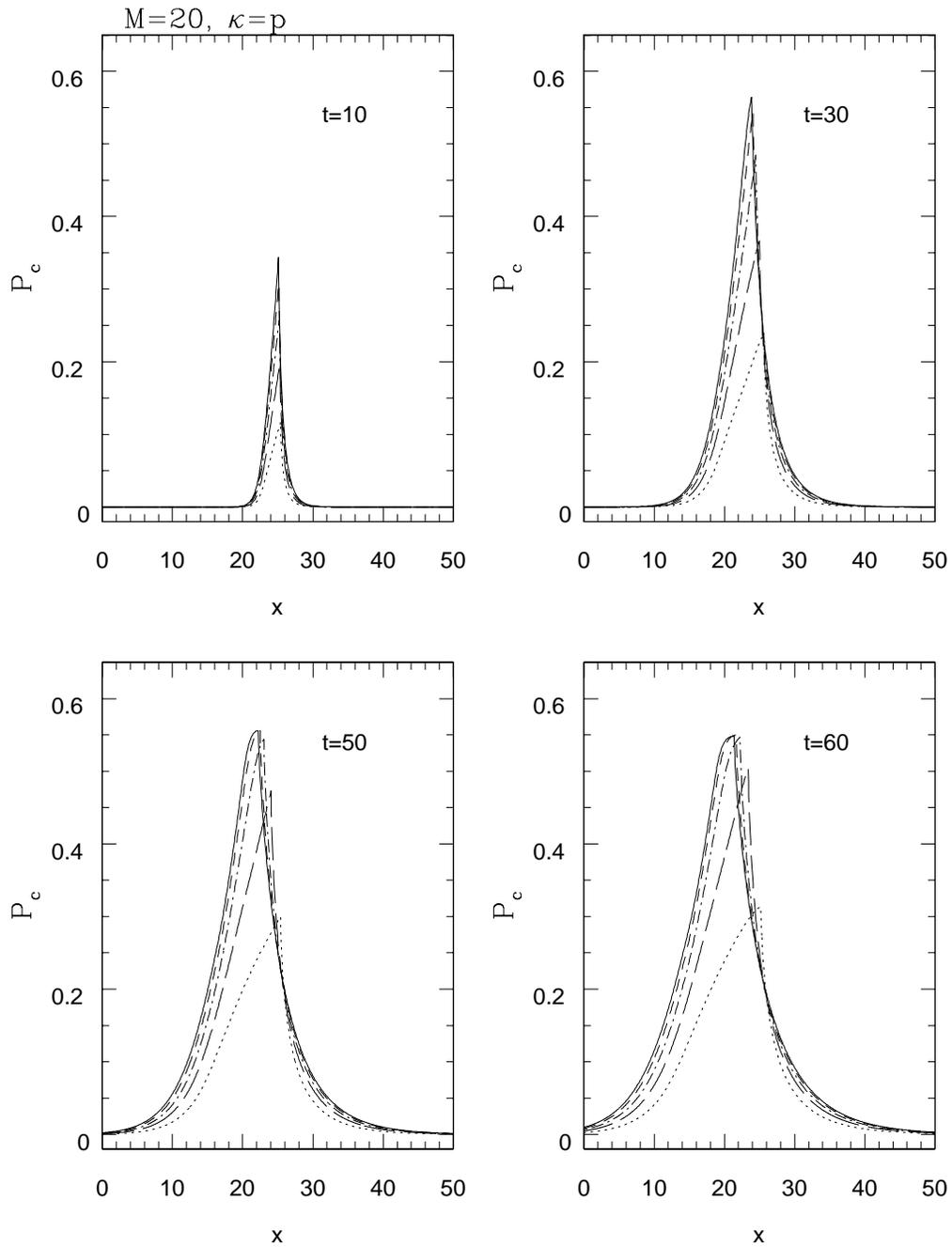}}
\figcaption{
Same as Fig. 4 except CR pressure is plotted.
\label{fig5}}
\end{figure}

\begin{figure}
\epsfxsize=14truecm
\centerline{\epsfbox{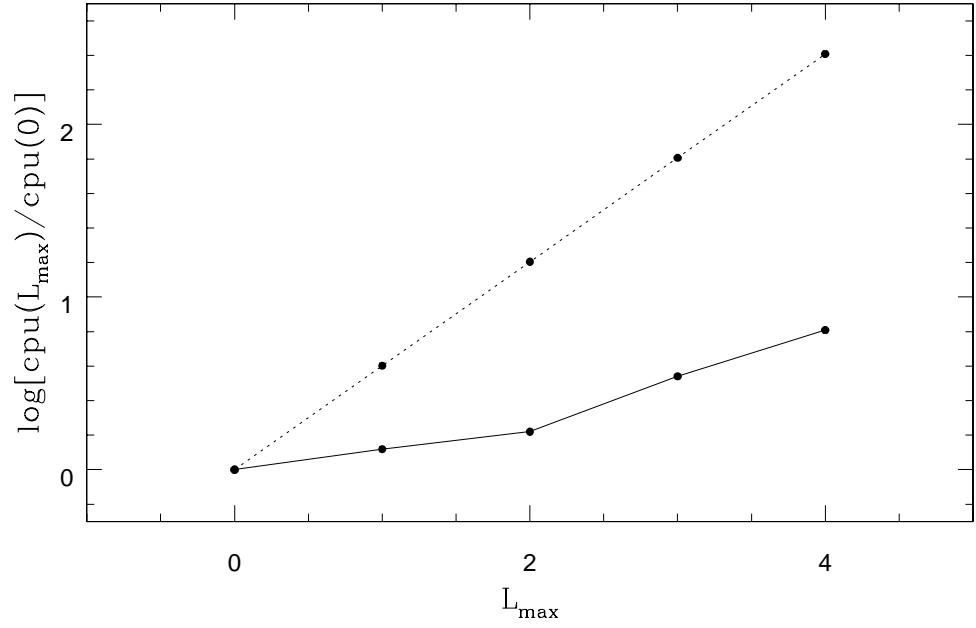}}
\figcaption{
The solid line shows the ratio of computational time required to 
include $L_{\rm max}$ levels relative to the time required with 
no refinement.
Here the number of cells at the base grid is $N = 2000$ and 
$N_{rf} = 200$ cells around the shock are refined.  
The dotted line shows the same ratio for the case when the finest 
resolution is applied over the entire grid, that is, 
$(2^{L_{\rm max}})^2$.  
\label{fig6}}
\end{figure}

\begin{figure}
\epsfxsize=14truecm
\centerline{\epsfbox{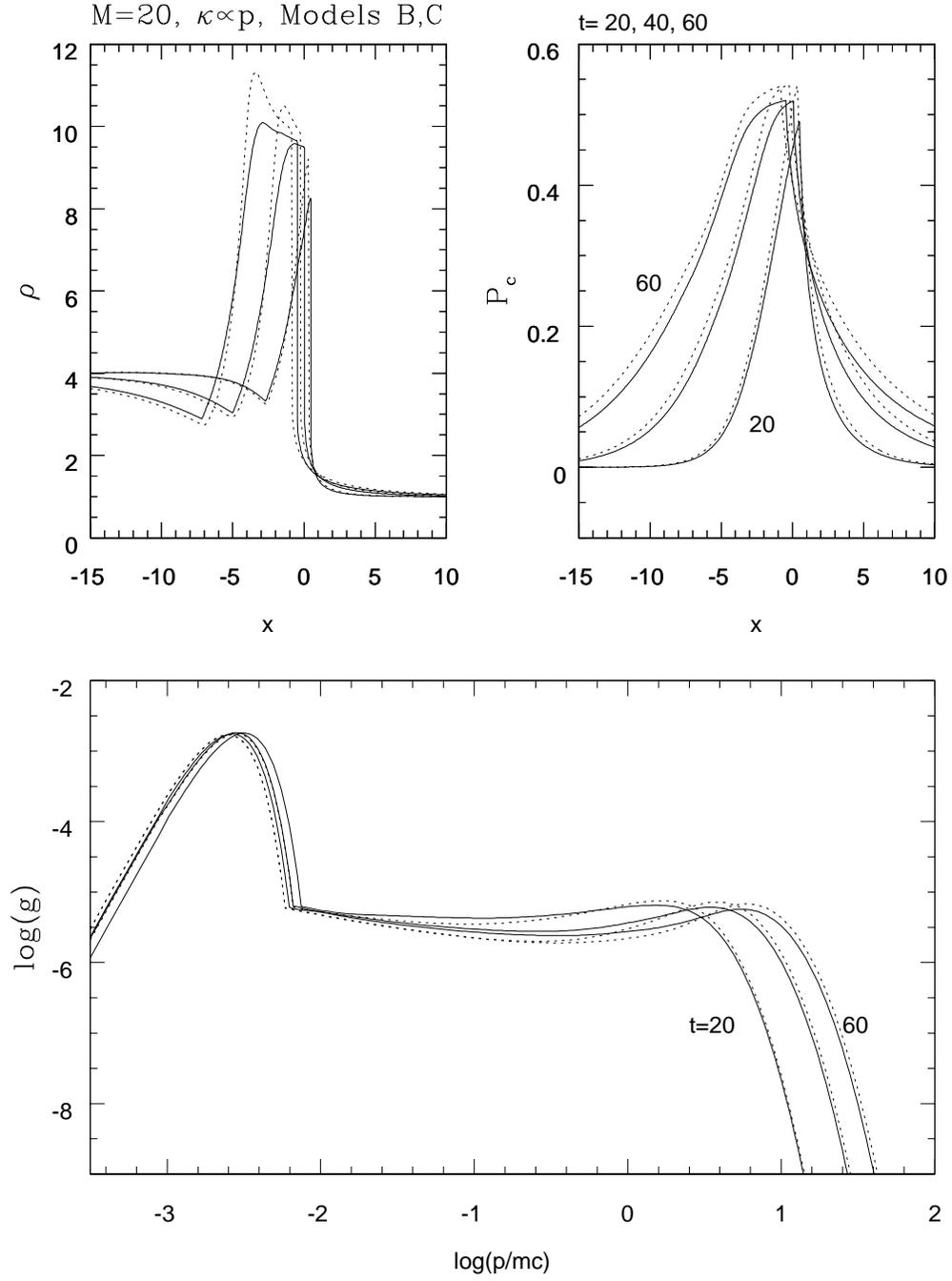}}
\figcaption{
Time evolution of the shock structure and 
the distribution function $g=fp^4$ at the shock
for Models B (dashed line) and C (solid line).
The shock moves to the left, so the right most plots correspond to the 
earliest time $t=20$ in the gas density and CR pressure plots. 
\label{fig7}}
\end{figure}

\begin{figure}
\epsfxsize=14truecm
\centerline{\epsfbox{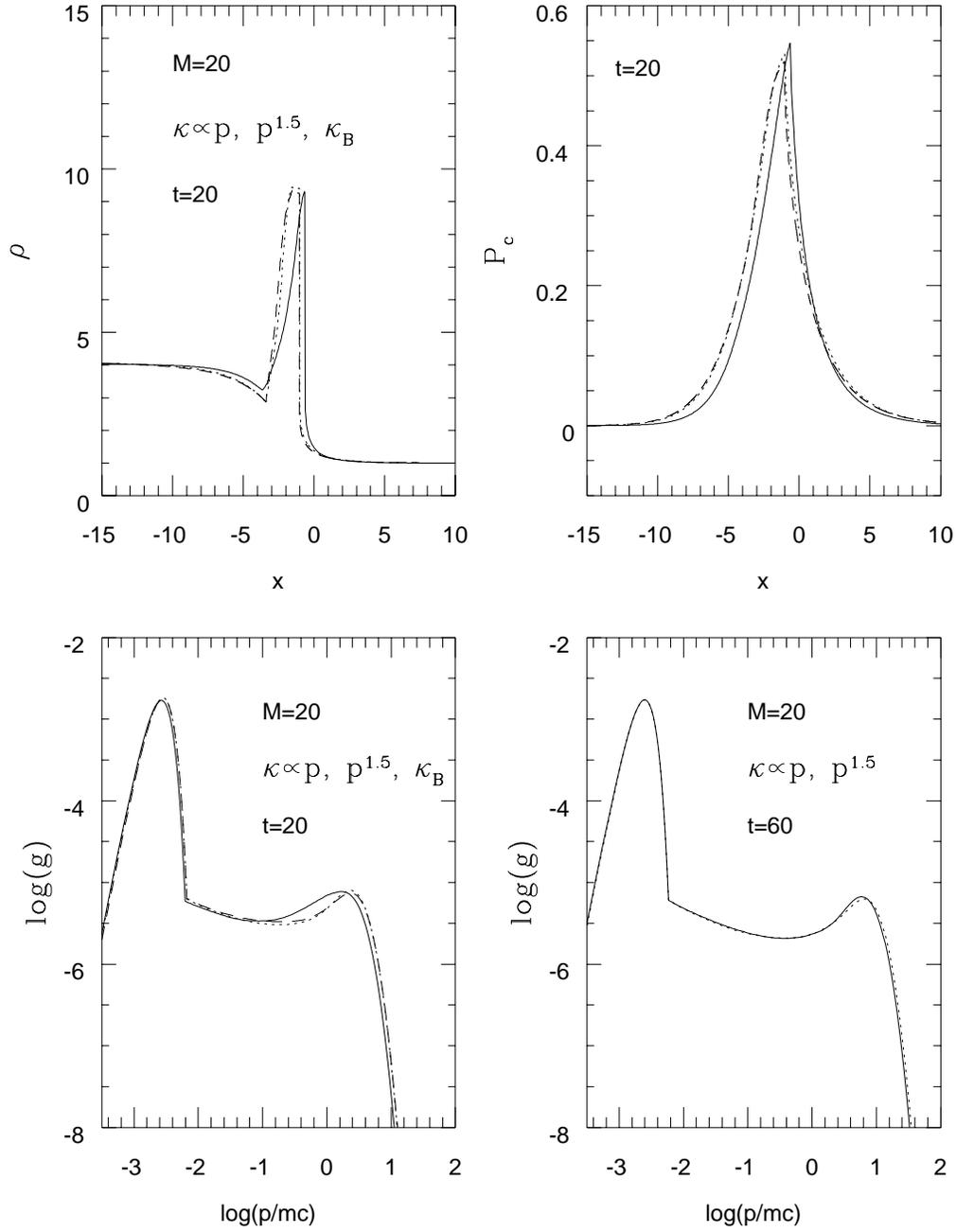}}
\figcaption{
Comparison of the $M=20$ shock structure at $t=20$ for Models K1 (solid), 
K2 (dotted), and K3 (dashed) with different diffusion coefficients (top
panels).
Comparison of the distribution function $g=fp^4$ at the shock
at $t=20$ for Models K1, K2, and K3 (bottom left) and
at $t=60$ for Models K1 and K2 (bottom right).
\label{fig8}}
\end{figure}

\begin{figure}
\epsfxsize=14truecm
\centerline{\epsfbox{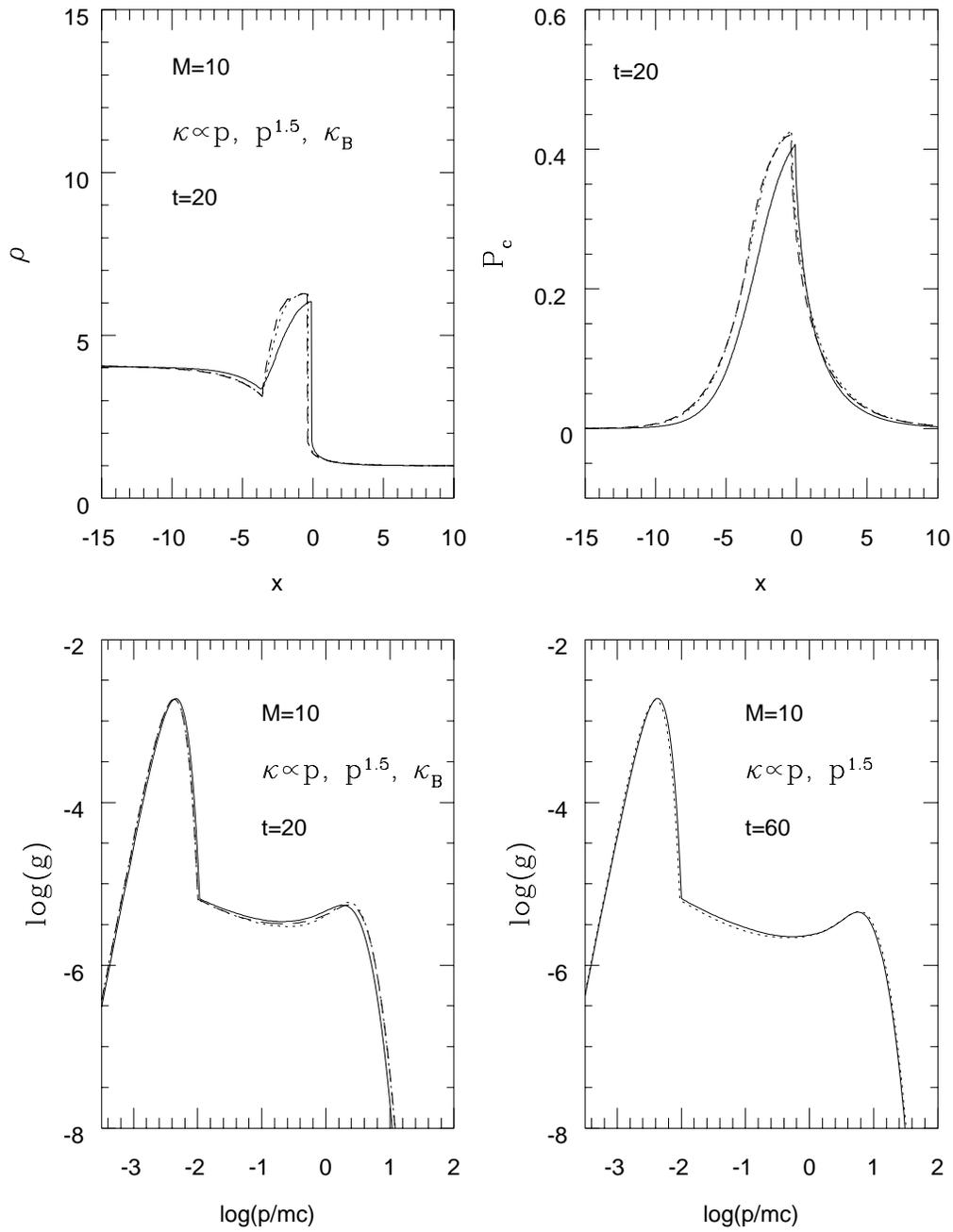}}
\figcaption{
Same as Fig. 8 except that the Mach number is 10.
\label{fig9}}
\end{figure}

\begin{figure}
\epsfxsize=14truecm
\centerline{\epsfbox{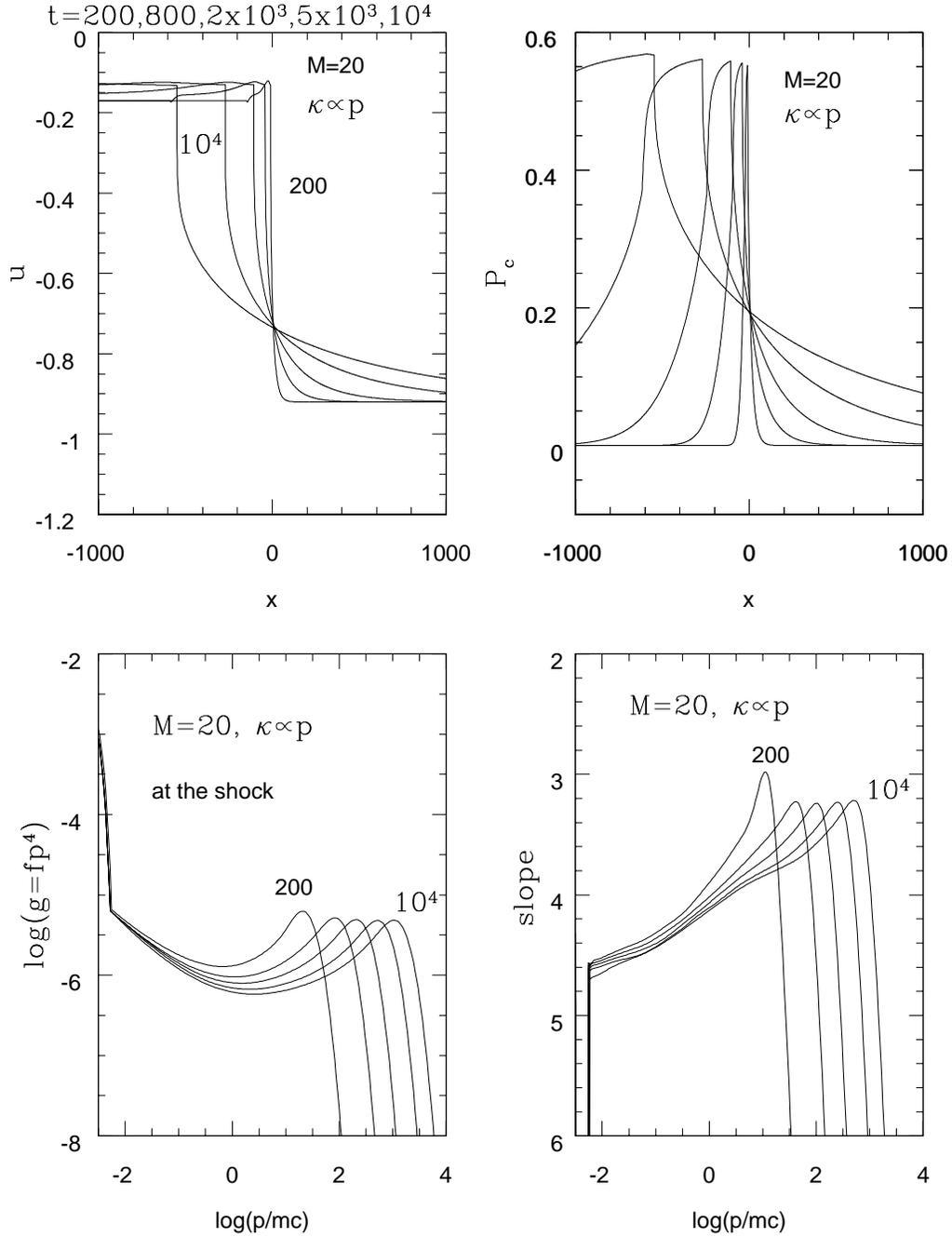}}
\figcaption{
Time evolution of the flow velocity and the CR pressure for 
the $M=20$ shock with $\kappa (p) \propto p$ for all momenta.
Also the CR distribution function $g$ at the shock and its power slope
$q=-\partial \ln{f} / \partial \ln{p}$.
The shock moves to the left, so the right most plots correspond to the 
earliest time $t=200$ in the flow velocity and CR pressure plots. 
\label{fig10}}
\end{figure}

\begin{figure}
\epsfxsize=14truecm
\centerline{\epsfbox{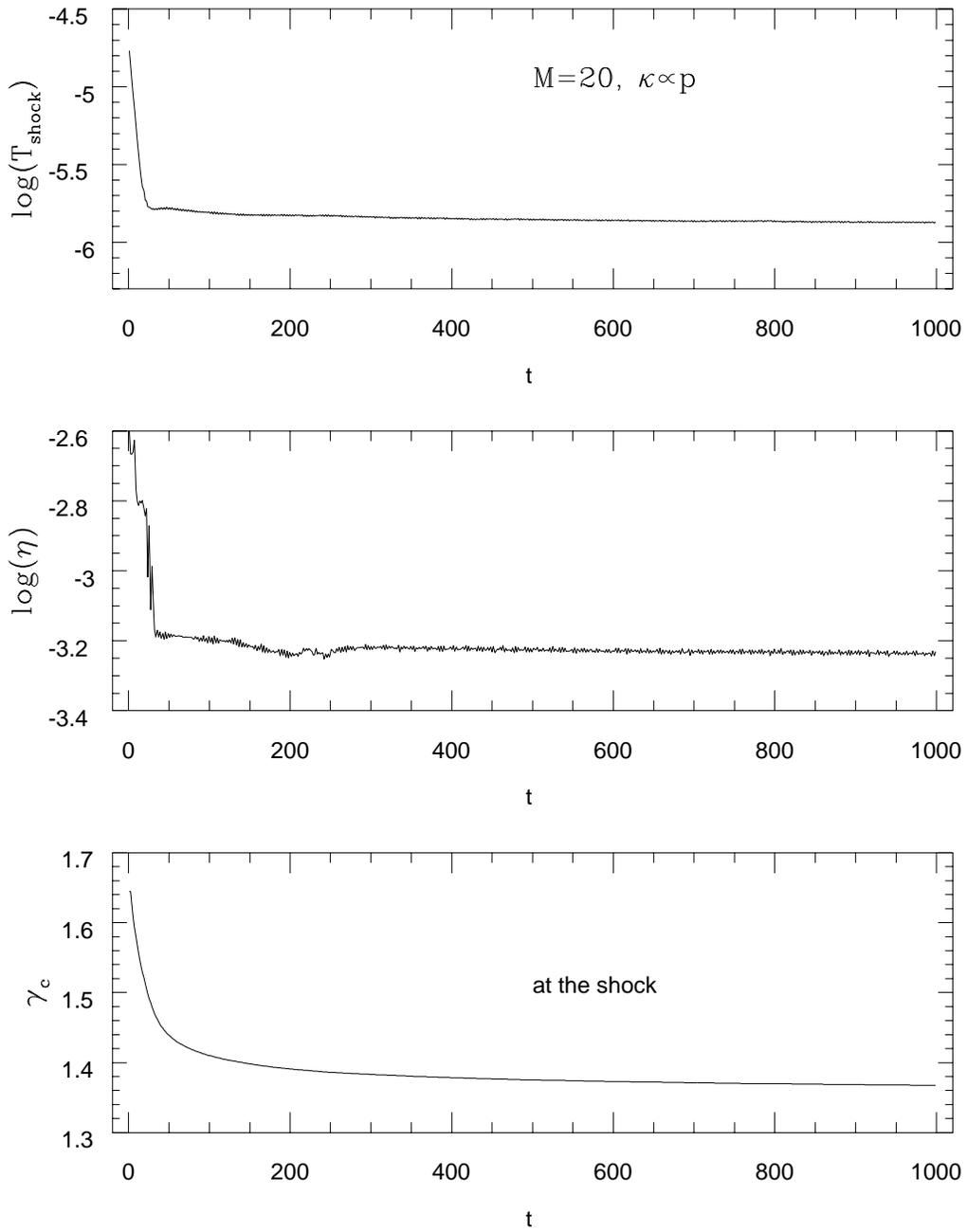}}
\figcaption{
Time evolution of the postshock shock temperature, injection rate, and
the adiabatic index of the CRs at the shock from the same simulation
shown in Fig. 10.  
\label{fig11}}
\end{figure}

\end{document}